\documentclass{nature}

\usepackage{graphicx}
\makeatletter
\let\saved@includegraphics\includegraphics
\AtBeginDocument{\let\includegraphics\saved@includegraphics}
\renewenvironment*{figure}{\@float{figure}}{\end@float}
\makeatother

\usepackage{bm}
\usepackage[version=3]{mhchem}
\usepackage{hyperref}
\usepackage{graphicx}
\usepackage{epsfig}
\usepackage{amsmath}
\usepackage{bbold}
\usepackage{float}
\usepackage{xcolor}



\title{A Quantum-Inspired Algorithm for the Factorized Form of Unitary Coupled Cluster Theory}

\author{Jia Chen$^{1,2}$, Hai-Ping Cheng$^{1,2}$ James Freericks$^3$}

\begin{document}

\maketitle
\begin{affiliations}
\item Department of Physics, University of Florida, Gainesville, FL 32611, USA
\item Quantum Theory Project,  University of Florida, Gainesville, FL 32611, USA

\item Department of Physics, Georgetown University, 37th St. and O St., NW, Washington, DC 20057, USA

\end{affiliations}

\begin{abstract}
 The factorized form of unitary coupled cluster theory (UCC) is a promising wave-function ansatz for the variational quantum eigensolver algorithm. Here, we present a quantum inspired algorithm for UCC based on an exact operator identity for the individual UCC factors. We implement this algorithm for calculations of the H$_{10}$ linear chain and the H$_2$O molecule with single and double $\zeta$ basis sets to provide insights about UCC as a wave-function ansatz. We find that for weakly-correlated molecules, factorized form UCC provides similar accuracy as conventional coupled cluster theory (CC); For strongly-correlated molecules where CC could break down, UCC significantly outperforms configuration interaction (CI) ansatz using the same excitations. As a result, factorized form UCC is a flexible and valuable electronic structure method, in both weakly and strongly correlated regions. UCC can also serve as initial state preparation method for the quantum phase-estimation algorithm, since it yields higher overlap with ground state than other common variational ansatze. 
\end{abstract}

\section{Introduction}

Unitary coupled cluster theory (UCC) was  proposed as a wave-function ansatz for quantum chemistry about four decades ago.\cite{Kutzelnigg1977, Koch1981, Kutzelnigg1983} In spite of efforts to develop computational methods for this ansatz, \cite{BARTLETT1989, WATTS1989, HOFFMANN1987, Taube2006} UCC is not nearly as prevalent as the closely related coupled cluster theory (CC) \cite{Bartlett2007, Shavitt2009} in electronic structure calculations. One major reason behind this lack of usage is because  the Baker-Campbell-Hausdorff expansion of the similarity-transformed Hamiltonian does not terminate in UCC (unlike what happens in CC), which often makes an exact calculation of this theory intractable on classical computers.

Quantum computing provides new ways to tackle the many-electron problem.\cite{Cao2019} For example, the quantum phase-estimation algorithm is capable of calculating the ground-state energy in polynomial time (provided a sufficiently good initial wavefunction is prepared, and time evolution is feasible for the Hamiltonian). Unfortunately, limitations on the maximal circuit depth that can be run on near-term quantum computers make the phase estimation algorithm unrealistic at this time. The variational quantum eigenvalues solver (VQE), \cite{Peruzzo2014} which is an approximate quantum-classical hybrid approach, has shown a lot of promise for the electronic structure problem in the near future. In VQE, a quantum circuit prepares a wavefunction that depends on a set of variational 
 parameters, and multiple measurement circuits are employed to then evaluate the expectation value of the energy (multiple circuits are needed because the Hamiltonian is broken up into a sum of unitary pieces to carry out the computation on a quantum computer). Optimization of the energy with respect to the variational parameters is then carried out in concert with a classical computer. Since most operations on a quantum computer are unitary, UCC has been proposed as a low-circuit-depth state-preparation ansatz for VQE. \cite{Romero_2018} 

This UCC-based state-preparation  strategy has been successfully implemented on an ion-trap quantum computers\cite{Hempel2018, Nam2020} for H$_2$O using a minimal basis and up to three UCC factors. This success further motivated research to improve both the UCC ansatz and the VQE algorithm; examples include, (i) the Bogoliubov transformed UCC \cite{Dallaire_Demers_2019} and  (ii) the k-UpCCGSD approach\cite{Lee2019} (which utilizes generalized single and double excitations). An adaptive algorithm has also been developed,\cite{Grimsley2019} which is able to adjust the ansatz by selecting the ordering of the most important operators to use. 

Due to the current paucity of robust quantum hardware, most UCC calculations are performed on classical computers. One way to do this is to directly simulate the quantum circuits on classical computers. This approach is limited, because quantum circuits with more than 50 qubits cannot be efficiently simulated. The other method available requires a numerical computation of the matrix exponential (via a truncated power series, rescaling and squaring, or diagonalization). Since the dimension of the matrices is equal to the dimensionality of the Hilbert space, only small systems can be studied this way.  As a result, the molecules that have been examined contain only a handful of atoms, and usually are represented by a minimal basis set. How the UCC will work for larger molecules and basis sets is critical to understanding how effective the VQE will be in advancing quantum chemistry on a quantum computer. Our work now allows us to do 
 this using classical computation.

In this paper, we introduce a quantum-inspired algorithm for a factorized form of UCC that is based on an operator identity  that recognizes a hidden $SU(2)$ symmetry.\cite{Evangelista2019,Xu2020} To illustrate how this approach works, we perform numerical calculations for the H$_{10}$ linear chain and for the H$_2$O molecule with the minimal and the double-$\zeta$ basis set. Analysis of these results sheds light onto the finer points of how one can employ the UCC on quantum computers.

\section{Theory and Method}

In UCC, the trial wave-function is expressed in an exponential form, given by
\begin{align}
    |\Psi_{\textrm{UCC}}\rangle = \exp (\hat{\sigma}) |\Psi_0\rangle , 
\end{align}
where $|\Psi_0\rangle $ is a single reference state and the operator $\hat{\sigma}$ is an anti-Hermitian combination of particle-hole excitation and de-excitation:
\begin{align}
    \hat{\sigma} &= \hat{T} - \hat{T}^{\dagger} ; \\
    \hat{T} &= \sum_i^{occ}\sum_a^{vir} \theta_i^a \hat{a}_a^{\dagger} \hat{a}_i +  \sum_{ij}^{occ}\sum_{ab}^{vir} \theta_{ij}^{ab} \hat{a}_{a}^{\dagger}\hat{a}_{b}^{\dagger} \hat{a}_{j}\hat{a}_{i} + \cdots ~,
\end{align}
where the angles $\theta$ are the variational parameters. We use letters from the start of the alphabet $a$, $b$, $c,\ldots$ to denote the virtual orbitals, with respect to the reference state,  and letters from the middle of the alphabet $i$, $j$, $k,\ldots$ to denote the occupied orbitals in the reference state. To simply notation, we express a general $n$-fold excitation operator as $\hat{a}^{ab\dots}_{ij\dots} = \hat{a}^{\dagger}_{a}\hat{a}^{\dagger}_{b} \dots \hat{a}_j \hat{a}_i$ (with the corresponding de-excitation operator being its Hermitian conjugate). We work in a factorized form for the  UCC, which is given by
\begin{align}
    |\Psi_{\textrm{UCC}}\rangle =\prod_{ij\cdots}^{occ}\prod_{ab\cdots}^{vir} \exp [\theta_{ij\cdots}^{ab\cdots} (\hat{a}_{ij\cdots}^{ab\cdots}-\hat{a}_{ab\cdots}^{ij\cdots})] |\Psi_0\rangle , 
    \label{eq:ucc-fac}
\end{align}
since this is the form of the UCC operator that can be easily generated on quantum computers. Note that this form is completely general, because we did not specify at all what the strategy is for determining the different factors. In particular, we can express the traditional UCC form (with all excitations appearing as a sum in the exponent) in this form, simply by using a Trotter breakup, which entails repeating many of the same factors in the expansion. 

Next, we discuss the operator identity for an arbitrary UCC factor appearing in Eq.~(\ref{eq:ucc-fac}) and use it to develop a classical algorithm for the UCC. 

\subsection{Operator Identity For UCC Factors}
We first examine UCC factors that correspond to single excitations:
\begin{align}
\hat{U}^a_i = \exp[\theta^a_i(\hat{a}^a_i - \hat{a}^i_a)].
\end{align}
This exponential operator can be reframed into the sum of a much simpler operator expression by using a hidden $SU$(2) group structure associated with these UCC factors. First define the ``hidden'' spin operators as
\begin{align}\label{singles}
\hat{S}_{+} = i\hat{a}^{\dagger}_a\hat{a}_i;\qquad \hat{S}_{-}=(\hat{S}_+)^{\dagger} = -i \hat{a}^+_i\hat{a}_a; \qquad \hat{S}_z = \frac{1}{2}[\hat{S}_+, \hat{S}_-]=\frac{1}{2}(\hat{a}^{\dagger}_a\hat{a}_a-\hat{a}^{\dagger}_i\hat{a}_i).
\end{align}
One can see that this is the conventional fermionic representation of spin, if we think of the virtual spin-orbital $a$ as corresponding to spin-up and the real spin-orbital $i$ as corresponding to spin down.
The commutation relations of these operators can then be immediately determined to be 
\begin{align}
[\hat{S}_z, \hat{S}_+] = 2\hat{S}_z, \qquad  [\hat{S}_z, \hat{S}_+]= \hat{S}_+, ~~\mathrm{and}~~~
[\hat{S}_z, \hat{S}_-] = -\hat{S}_- ,
\label{eq:su2_alg}
\end{align}
which can be recognized as the conventional $SU$(2) algebra. 
But these operators are not an independent $SU$(2) algebra, instead they arise as a subgroup of the permutation symmetry of all the generators of the allowed UCC factors (in factorized form). Hence, they represent a direct sum of $S=0$ and $S=1/2$ representations when acting on any product state in the Hilbert space. We see this when we examine some additional operator identities given by 
\begin{align}\label{spin_operator_identity}
\begin{split}
\hat{S}_+^2 &= \hat{S}_-^2=0; ~~(\hat{S}_+ + \hat{S}_-)^2 = \hat{S}_+\hat{S}_-+\hat{S}_-\hat{S}_+; \\ (\hat{S}_+ + \hat{S}_-)^3 &= \hat{S}_+\hat{S}_-\hat{S}_+ + \hat{S}_-\hat{S}_+\hat{S}_- =  2\hat{S}_z\hat{S}_+ - 2\hat{S}_z\hat{S}_- = \hat{S}_+ + \hat{S}_- ,
\end{split}
\end{align}
which are not general operator identities of the $SU$(2) algebra, but are specific to this direct-sum space.
These identities immediately imply that odd powers of $\hat{S}_+ + \hat{S}_-$ are equal to $\hat{S}_+ + \hat{S}_-$ and even powers are equal to $\hat{S}_+\hat{S}_-+\hat{S}_-\hat{S}_+$. These identities are similar to a spin-1 representation, where the cube of the Cartesian angular momentum operators are equal to the Cartesian angular momentum operators. Indeed, this same identity allows us to evaluate the exponentials exactly. Simply expand the exponential of the corresponding UCC factor in a power series and use the fact that all nonzero even powers of the operator are the same and all odd powers  of the operator are the same. Then the numerical factors can be immediately resummed to yield
\begin{align}\label{power_series}
\begin{split}
\exp[-i\theta(\hat{S}_++\hat{S}_-)] &= 1 - i\theta(\hat{S}_+ + \hat{S}_-) + \frac{(i\theta)^2}{2}(\hat{S}_+ + \hat{S}_-)^2 - \frac{(i\theta)^3}{3!}(\hat{S}_+ + \hat{S}_-)^3+\cdots \\
&= 1 - i\sin\theta(\hat{S}_+ + \hat{S}_-) + (\cos\theta-1)(\hat{S}_+\hat{S}_-+\hat{S}_-\hat{S}_+).
\end{split}
\end{align}
For this single UCC excitation, inserting Eq.~(\ref{singles}) into Eq.~(\ref{power_series}) yields:
\begin{align}\label{single_identity}
\exp[\theta (\hat{a}^a_i-\hat{a}^i_a)] = 1 + \sin\theta(\hat{a}^a_i-\hat{a}^i_a) + (\cos\theta -1)(\hat{n}_a + \hat{n}_i - 2\hat{n}_a\hat{n}_i),
\end{align}
where $\hat{n}_\alpha = \hat{a}_\alpha^{\dagger}\hat{a}_\alpha$ is the density operator for $\alpha=a$ or $i$. The result of applying Eq.~(\ref{single_identity}) onto certain states can be found in Ref.~\citen{Evangelista2019}. More importantly, this operator identity can be generalized to any UCC factor of arbitrary $n$-particle excitations via the recognition of the hidden $SU$(2) algebra for the general case, which is given by
\begin{align}
\begin{split}
&\hat{S}_+ = i\hat{a}^{a_1a_2\cdots a_n}_{i_1i_2\cdots i_n}, ~~ \hat{S}_- = -i \hat{a}^{i_1i_2\cdots i_n}_{a_1a_2\cdots a_n},~~\mathrm{and} \\ &\hat{S}_z=  \frac{1}{2} \left (\hat{n}_{a_1}\dots\hat{n}_{a_n}(1-\hat{n}_{i_1})\cdots(1-\hat{n}_{i_n})
- (1-\hat{n}_{a_1})\dots(1-\hat{n}_{a_n})\hat{n}_{i_1}\cdots\hat{n}_{i_n} \right ).
\end{split}
\end{align}
These spin operators satisfy the $SU$(2) commutation relations in Eq.~(\ref{eq:su2_alg}) and also satisfy the additional operator identities in
Eq.~(\ref{spin_operator_identity}). Hence, we can perform the exact same expansion of the power series for this UCC factor and find the same result as given in Eq.~(\ref{power_series}). Evaluating the spin operators in terms of the excitation and de-excitation operators then yields the final exact operator identity:
\begin{align}\label{operator_identity}
\begin{split}
& U^{a_1\dots a_n}_{i_1\dots i_n} = \exp[\theta(\hat{a}^{a_1\dots a_n}_{i_1\dots i_n}-\hat{a}^{i_1\dots i_n}_{a_1\dots a_n})] = 1 + \sin\theta (\hat{a}^{a_1\dots a_n}_{i_1\dots i_n}-\hat{a}^{i_1\dots i_n}_{a_1\dots a_n}) \\
&+(\cos\theta-1)[\hat{n}_{a_1}\dots\hat{n}_{a_n}(1-\hat{n}_{i_1})\dots(1-\hat{n}_{i_n}) +(1-\hat{n}_{a_1})\dots(1-\hat{n}_{a_n})\hat{n}_{i_1}\dots\hat{n}_{i_n}].
\end{split}
\end{align}
This identity acts in a direct sum space of $S=0$ and $S=1/2$: when $S=0$, which happens when no excitation or de-excitation is possible, the operator acts as the identity, but when $S=1/2$, which happens when an excitation or de-excitation is possible, the operator acts analogous to a spin-one-half spinor, which is rotated by the angle $\theta$.

\subsection{The Quantum-Inspired Algorithm}
Based on the exact operator identity derived above, we devise an algorithm inspired by the VQE \cite{Peruzzo2014} for UCC in a factorized form; this algorithm can be carried out on classical computers. Eq.~(\ref{operator_identity}) guarantees that applying the UCC factor $U^{a_1\cdots a_n}_{i_1\dots i_n}$ to the configuration $ |\Psi^{a_{1'}\cdots a_{n'}}_{i_{1'}\cdots i_{n'}}\rangle$, where $ |\Psi^{a_{1'}\cdots a_{n'}}_{i_{1'}\cdots i_{n'}}\rangle$ is the result of applying $\hat{a}^{a_{1'}\cdots a_{n'}}_{i_{1'}\cdots i_{n'}}$ to reference state $|\Psi_0\rangle$, can only have three outcomes: 
\begin{enumerate}
	\item ($S=1/2$ case, excitation) If sets $\{a_1\cdots a_n\}$ and $\{a_{1'}\cdots a_{n'}\}$ have no common elements; and sets $\{i_1\cdots i_n\}$ and $\{i_{1'}\cdots i_{n'}\}$ have no common elements, then applying the UCC factor generates the sum of two configurations---the original configuration $ |\Psi^{a_{1'}\cdots a_{n'}}_{i_{1'}\cdots i_{n'}}\rangle$ with coefficient $\cos\theta$ and the excitation $ |\Psi^{a_{1'}\cdots a_{n'} a_{i}\cdots a_{n}}_{i_{1'}\cdots i_{n'}i_{1}\cdots i_{n}}\rangle$ with coefficient $\sin\theta$.
	\item ($S=1/2$ case, de-excitation) If set $\{a_1\cdots a_n\}$ is subset of $\{a_{1'}\cdots a_{n'}\}$ with complement set $\{ a_{1''}\cdots a_{n''}\} = \{ x \in \{a_{1'}\cdots a_{n'}\} | x \notin \{a_1\cdots a_n\}  \}$;
	and set $\{i_1\cdots i_n\}$ is a subset of $\{i_{1'}\cdots i_{n'}\}$ with complement set $\{ i_{1''}\cdots i_{n''}\} = \{ x \in \{i_{1'}\cdots i_{n'}\} | x \notin \{i_1\cdots i_n\}  \}$, then applying UCC factors generates two configurations---the original $|\Psi^{a_{1'}\cdots a_{n'}}_{i_{1'}\cdots i_{n'}}\rangle$ with coefficient $\cos\theta$ and the de-excitation $|\Psi^{a_{1''}\cdots a_{n''} }_{i_{1''}\cdots i_{n''} }\rangle$ with coefficient $-\sin\theta$.
	\item ($S=0$ case, nothing) If the above two conditions are not satisfied, then the result is one configuration, the original $|\Psi^{a_{1'}\cdots a_{n'}}_{i_{1'}\cdots i_{n'}}\rangle$ with coefficient 1. 
\end{enumerate}

\begin{figure}
\includegraphics[width=\linewidth,clip]{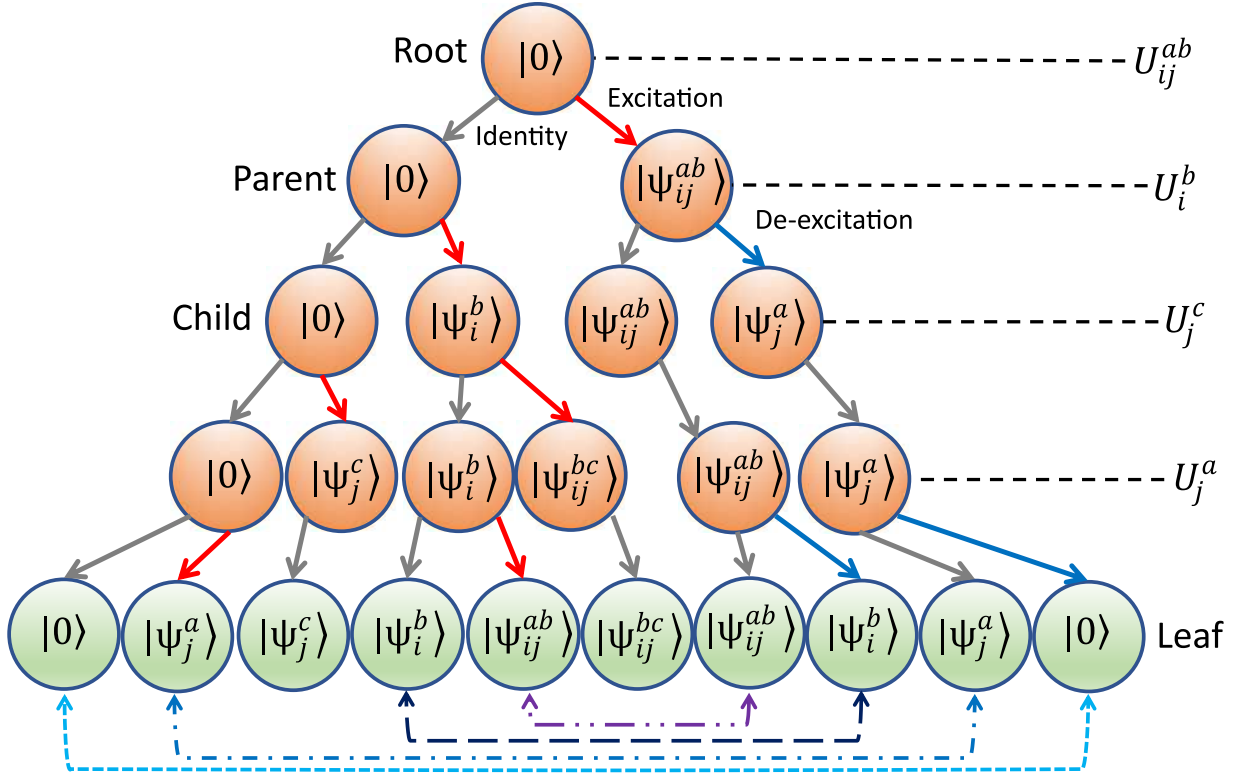}
\caption{ Tree structure of configurations generated by three UCC factors. Red arrows represent excitation, grey identity and blue de-excitation. Lines with two arrows at the bottom level represent elimination of the repeating configurations.} \label{tree}
\end{figure}

Equipped with these exact rules, we can efficiently produce UCC wavefunctions (in the factorized form) on classical computers. Configurations generated by sequentially applying UCC factors fit into a tree-type data structure. The root node of the tree is the initial reference configuration. Each level except the last one corresponds to the application of one of the UCC factors in sequential order. Each parent node gives rise to one or two child nodes according to Eq.~(\ref{operator_identity}) and the three rules stated above. Starting from the root node with coefficient 1, we can calculate the coefficients for all the nodes on the tree according to the exact rules. The last level of the tree (leaf nodes) is the final UCC wave-function expressed as a linear combination of configurations. It is important to condense all final leaves (those indicated by the interrupted line pairs in the last row of the figure) to save space during the tree construction.  One example of this tree structure built on three UCC factors can be found in Fig.~\ref{tree}.

This tree structure can not only be used to generate UCC wave-function, but also the derivatives of the wavefunction with respect to the variational parameters, which is extemely useful in optimizing the variational ansatz. We can take the derivative of Eq.~(\ref{operator_identity}):
\begin{align}\label{operator_identity_2}
\begin{split}
&\frac{d \hat{U}^{a_1\dots a_n}_{i_1\dots i_n} (\theta)}{d \theta} =  \cos\theta (\hat{a}^{a_1\dots a_n}_{i_1\dots i_n}-\hat{a}^{i_1\dots i_n}_{a_1\dots a_n}) \\
&-\sin\theta [\hat{n}_{a_1}\dots\hat{n}_{a_n}(1-\hat{n}_{i_1})\dots(1-\hat{n}_{i_n}) +(1-\hat{n}_{a_1})\dots(1-\hat{n}_{a_n})\hat{n}_{i_1}\dots\hat{n}_{i_n}].
\end{split}
\end{align}
The result of applying this operator to a configuration is similar to what has been discussed above, but with coefficients modified. The derivative of the wavefunction with respect to each variational parameter immediately follows as
\begin{align}
   \frac{ d |\Psi_{\textrm{UCC}}\rangle}{d \theta_i} = \hat{U}_n\cdots\hat{U}_{i+1}\frac{d \hat{U}_i}{d \theta_i}\cdots\hat{U}_2\hat{U}_1|\Psi_0\rangle .
\end{align}

The Hamiltonian of the molecule is constructed from the reference state. We use canonicalized Hartree-Fock (HF) wave-functions as the reference state. Second quantized Hamiltonian becomes
\begin{align}
    \hat{\mathcal H} = h_0 + \sum_{pq}h_{pq}\hat{a}^{\dagger}_p\hat{a}_q + \frac{1}{2}\sum_{pqrs} \hat{a}^{\dagger}_p\hat{a}^{\dagger}_q\hat{a}_r\hat{a}_s,
\end{align}
where $p,q,r,s$ label orthogonal spin-orbitals; $h_0$ is the nuclear repulsion energy, and $h_{pq}$ and $h_{pqrs}$ are one- and two-electron integrals generated by \textbf{PySCF} package:\cite{Sun2018, Sun2020}
\begin{align}
    h_{pq} = \int dr \phi_p(r) \left( -\frac{\nabla}{2} - \sum_i\frac{Z_i}{|R_i - r|}\right) \phi_q(r);
\end{align}

\begin{align}
    h_{pqrs} = \int dr_1 dr_2\frac{\phi_p(r_1)\phi_q(r_2)\phi_s(r_1)\phi_r(r_2)}{|r_1 - r_2|} .
\end{align}

 To initialize these calculations, we need a set of UCC factors in a particular order. Different orderings of the same set of factors can possibly correspond to different wavefunction ansatzes, as pointed out previously\cite{Grimsley2020} (and verified by simply looking carefully at how the tree is formed).  In this work,  we use second-order Møller–Plesset perturbation theory \cite{MP1934} (MP2) to choose the UCC factors and their ordering. MP2 provides amplitudes for double excitations, which are easy to obtain and serve as good estimations of their importance. UCC factors are chosen in descending order of the absolute value of the corresponding MP2 amplitudes. This strategy to order UCC factors is well defined and provides a concrete selection and ordering scheme for the UCC factors, as the later factors are assumed to be less relevant. A drawback is that MP2 only has amplitudes for double excitations; to include single excitations in calculations, we put the single factors after the double factors, with random order. As we will see later, variational parameters associated with single UCC factors are generally small, thus their ordering ends up being of minor significance. Non-HF starting point, like natural orbitals could be used to improve screening of UCC factors in the future.

\begin{figure}
\includegraphics[width=0.6\linewidth,clip]{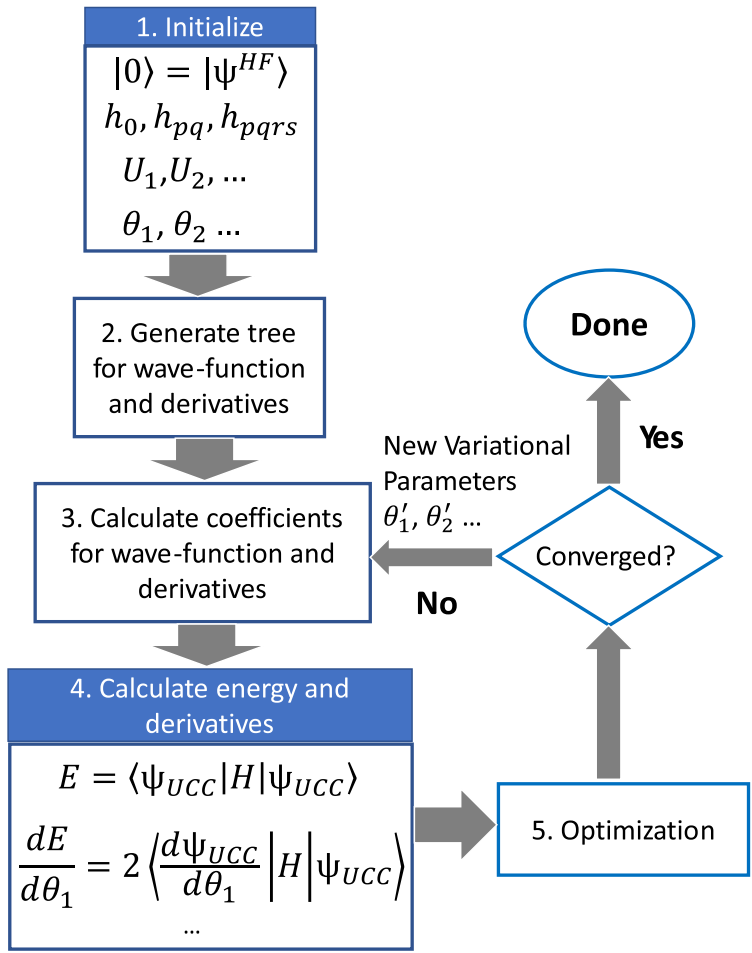}
\caption{Flowchart of the quantum-inspired algorithm for factorized form of UCC. } \label{flow}
\end{figure}

 With Hartree-Fock spin-orbitals, a parameterized Hamiltonian, UCC factors chosen in MP2 order, and MP2 amplitudes as the initial guess for variational parameters, the initialization step is complete. In the next step, the tree structure of the configurations is generated. When the number of UCC factors is large, the number of configurations on the tree becomes prohibitively large. Eliminating repeating configurations on each level of the tree greatly reduces the memory requirement of the calculation. But this also means one child node can have more than one parent node. In the third step, coefficients for all the configurations on the tree are calculated; the coefficients for the wavefunction and its derivatives are obtained from the leaf nodes. Then, the energy and its derivatives are evaluated from the expectation values of Hamiltonian. Subsequently, the energy and derivatives are fed into an optimization algorithm. We used the Broyden–Fletcher–Goldfarb–Shannon (BFGS) minimization scheme as implemented in \textbf{SciPy}.\cite{2020SciPy-NMeth} If convergence is not achieved, an updated set of values for the variational parameters are employed to recalculate the coefficients. Since the tree structure is fixed and saved in memory, it is not necessary to re-generate it during optimization; this procedure greatly saves time in completing the calculation. This then is the classical algorithm for UCC in factorized form, and its flowchart is illustrated in Fig.~\ref{flow}. To speed up the energy and derivative calculations, we prune the leaf nodes and keep only configurations with an absolute value of their amplitude larger than a specified threshold. In this work, we set the threshold to $10^{-6}$, and differences in total energies due to pruning are always found to be smaller than $10^{-5}$~Ha in several test cases. 

\section{Results and Discussion}
\subsection{H$_{10}$ Linear Chain}
\begin{figure}
\includegraphics[width=0.8\linewidth,clip]{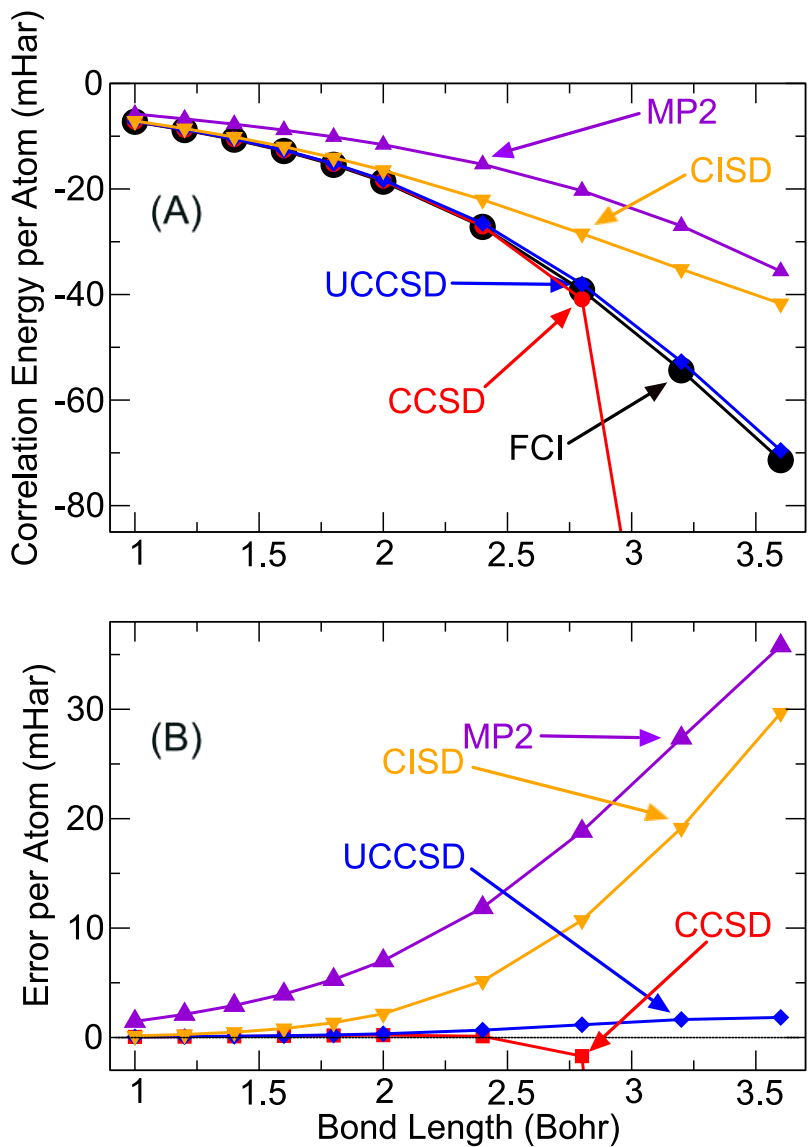}
\caption{Correlation energy (panel A) and error (panel B) calculated with several quantum chemistry methods as a function of the bond length of the hydrogen linear chain. Data for CCSD are from Ref.~\citenum{Motta2017}. Data for other methods are from calculations performed in this work.}  \label{H10_energy}
\end{figure}

We apply this quantum inspired UCC algorithm  to calculate the ground state of the H$_{10}$ linear chain with the minimal basis set STO-6G. We choose this model system because extensive benchmark calculations using many state-of-art computational methods have already been applied to it.\cite{Motta2017} Comparison of our results with these standards provides insight into the accuracy of UCC as an electronic structure tool. 
With the minimal basis set, the reference state of the H$_{10}$ linear chain has 10 molecular orbitals, and half of them are occupied. The Hilbert space for the $\langle \hat{s}_z\rangle=0$ sector contains  63,504 determinants, with 825 of them doubles and 50 of them singles; we use a lower case $\hat{s}_z$ to refer to the physical $z$-component of spin of the different product states. Correlation energies (difference between the calculated energy and the Hartree-Fock energy) and the error (difference between the calculated energy and the full configuration interaction  (FCI) energy) of several methods as a function of bond length between the H atoms are plotted in Fig.~\ref{H10_energy}.

The correlation energy of the hydrogen chain increases with the bond length, which indicates that the system crosses over from being weakly correlated to being strongly correlated. In the weakly correlated regime, UCCSD results are very close those obtained by CCSD, which is known to be the method of choice for weakly correlated molecules. When the correlation strength goes beyond a certain point (where the bond length is approximately 2.5 Bohr in this case), CCSD energy is no longer bounded by FCI from below, and the error can become quite large. As constructed, the UCCSD calculation is always variational. Calculations showed UCCSD curve followed FCI results closely. Compared to the configuration interaction singles and doubles (CISD), which is also variational, and the MP2 method, which is not, the UCCSD provides significantly improved energies for all the bond lengths covered here. 

\begin{figure}
\includegraphics[width=0.8\linewidth,clip]{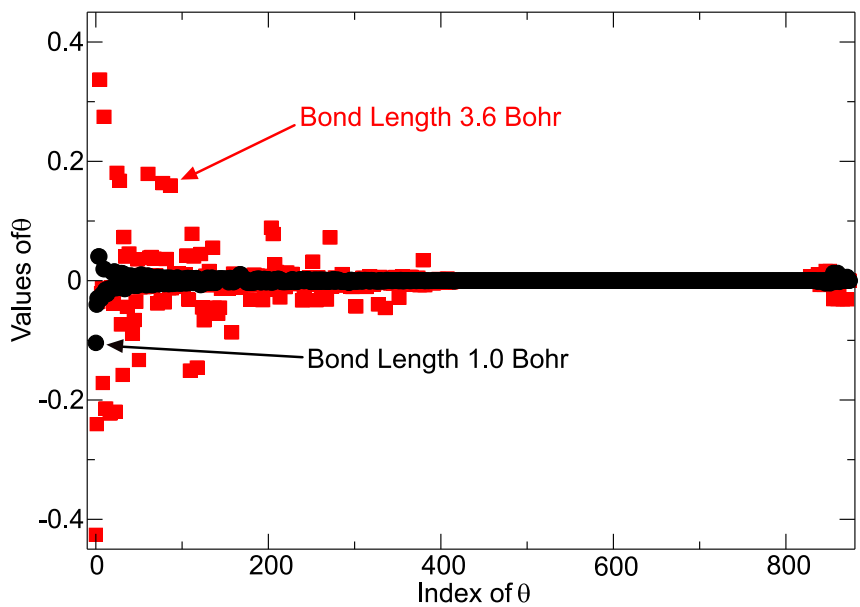}
\caption{ Variational parameters $\theta$ obtained from energy optimization at bond lengths of 1.0 (black circles) and 3.6 Bohr (red squares). The first 825 $\theta$ values are for UCC double factors and last 50 are for the UCC singles factors. } \label{angles}
\end{figure}

Additional insights about the UCC can be obtained by examining the final variational wavefunctions generated in the calculations. Values of the variational parameters $\theta$ after optimization, for the weakly and strongly correlated systems are plotted in Fig.~\ref{angles}. A couple of observations can be made about the values of $\theta$. First, a large number of UCC factors of double excitations have $\theta$ very close to zero (here essentially all  UCC doubles factors with index larger than 450; recall the singles come last). Those factors can be ignored in the UCC calculations if we have good estimations of each UCC factor's importance beforehand. Second, the absolute values of $\theta$ are tied to correlation energies. Strongly correlated systems (such as the bond length of 3.6 Bohr, plotted with red squares) have much larger $\theta$ values than the weakly correlated system (such as the bond length of 1.0 Bohr, plotted with black circles). It is interesting that all of these angles, even in strongly correlated systems, are clustered close to $\theta=0$.

\begin{figure}
\includegraphics[width=0.8\linewidth,clip]{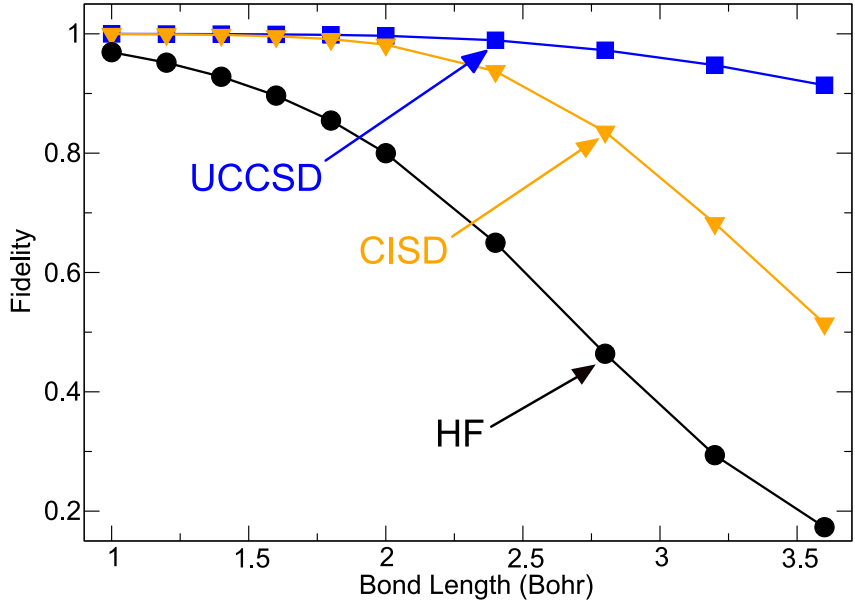}
\caption{ Fidelity of wavefunction prepared by HF, CISD and UCCSD, which are calculated as the squared overlap between the approximate wavefunction $|\psi_{\rm approx.}\rangle$ and exact ground-state wavefunction (from FCI calculations): $F = |\langle \psi_{\rm approx.} | \psi_{FCI} \rangle |^2$  } \label{fide}
\end{figure}

The quantum phase-estimation algorithm is the ideal approach to calculate the ground-state energy (and to prepare the ground state) on a quantum computer; it should be possible to use this method once fault-tolerant quantum computation with high depth circuits is feasible.  One potential application of the factorized form of UCC is to use it as the initial wave-function preparation method to start the phase-estimation algorithm. The success rates of the phase-estimation algorithm depend on the fidelity of the initially prepared wave-functions, which is determined by the squared overlap between the prepared wave-function and the exact ground state wave-function, (FCI wave-function): $F = |\langle \psi_{\rm approx.} | \psi_{\rm FCI} \rangle |^2$. Since, HF and CISD are variational, their states can in principle be prepared on quantum computers, with the Hartree-Fock being trivial in a second-quantized formalism. The fidelity of these approximate wavefunctions is plotted in Fig.~\ref{fide}.  We see that factorized form of UCCSD provides a wavefunction with higher fidelity for all bond lengths studied here. The difference is significant in the strongly-correlated regime, which makes UCC a much more suitable state preparation method for the phase-estimation algorithm in those cases. 

\subsection{H$_2$O with single and double $\zeta$ basis sets}
\begin{figure}
\includegraphics[width=0.8\linewidth,clip]{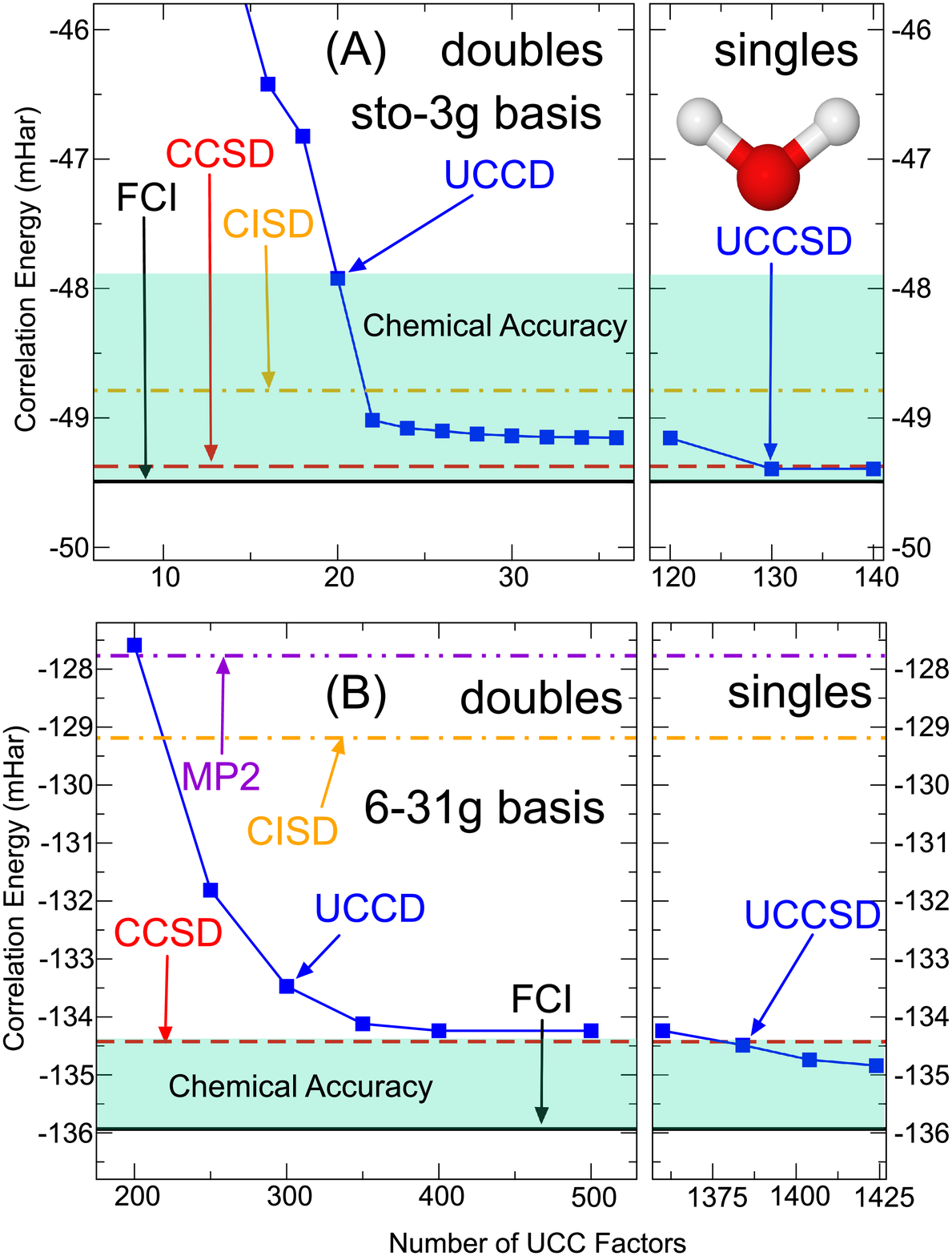}
\caption{  Panel(A): correlation energy of the H$_2$O molecule from a UCC calculation with the single-$\zeta$ basis set STO-3g. The first 120 UCC factors are doubles, and the last 20 are singles. Panel (B): correlation energy from a calculation with the double-$\zeta$ basis set 6-31g. The first 1360 UCC factors are doubles and last 64 are singles. } \label{H2O}
\end{figure}

The next molecule we study is the H$_2$O molecule. With a minimum basis set STO-3g, a calculation using 3 UCC factors has been performed on a trapped ion quantum computer, and simulations (with more factors) on a classical computer. These latter calculations showed that only a small number of UCC factors (about 20) are necessary for UCC to achieve chemical accuracy, which is an energy window of 1.6 milliHartree (mHar) from the ground state.\cite{Nam2020} With the STO-3g basis set, the H$_2$O molecule has 7 molecular orbitals and 10 electrons, which leads to 120 double excitations and 20 single excitations. We put double and single UCC factors in descending order of their MP2 amplitudes, and applied our algorithm to this system. Similarly, we find that 20 UCC factors, (all doubles in our case) are needed to reach chemical accuracy (see Fig.~\ref{H2O}). After incorporating all the double and single UCC factors, the UCCSD energies are slightly better than CCSD, and they are much better than CISD. 

It is known that the performance of the minimum basis sets is generally poor, and a double-$\zeta$ basis set can give a striking improvement.  \cite{big_purple_book} Here, we perform a UCC calculation for the H$_2$O molecule with a 6-31g basis set, which has 13 molecular orbitals. To make the calculations more manageable, the orbital of lowest energy (oxygen 1s orbital) is frozen. In total, this system (with one frozen orbital) has 1360 double and 64 single excitations. Results of these calculations are plotted in panel (B) of Fig.~\ref{H2O}. First, similar to the H$_{10}$ linear chain, we find that after about 400 doubles, the remaining 1000 doubles UCC factors do not further improve the energy. The energy after including all the single and double UCC factors is again, slightly lower than CCSD, and much better than MP2 and CISD. Unlike the system with sto-3g basis set, the chemical accuracy can only be achieved after including all the important singles and doubles, which indicates chemical accuracy is harder to reach when we use larger basis sets. 

\section{Conclusion}
In this work, we presented a classical algorithm for the factorized form of UCC, which was inspired by the quantum VQE algorithm. The foundation of this algorithm is an operator identity for the general UCC factor, uncovered from a hidden $SU(2)$ symmetry. Implementation of this algorithm allowed us to apply UCC to larger systems than studied before on classical computers. Compared to more established quantum chemistry methods, factorized form UCC is as accurate as CC for weakly correlated systems, and provides much better results than CI with the same excitations for strongly correlated systems. Since we can choose factors and their orders, factorized form UCC can be molded for different needs and computational budgets. The accuracy and flexibility make UCC a very valuable tool for chemistry, material science and condensed matter physics. 
 
\section{Data availability}
The data that support the findings of this study are available from the corresponding author upon reasonable request

\begin{addendum}
 \item JC and HPC are supported by the Department of Energy, Basic Energy Sciences under contract DE-FG02-02ER45995. JKF is supported from the National Science Foundation under grant number CHE-1836497. JKF is also funded by the McDevitt bequest at Georgetown University. We also acknowledge useful discussions with Rodney Bartlett, Joseph Lee, John Staunton, Cyrus Umrigar, Luogen Xu and Dominika Zgid.

 \item[Competing Interests] The authors declare that they have no
competing financial interests.
\item[Correspondence] Correspondence and requests for materials should be addressed to H-P. C. (email: hping@ufl.edu) and J. K. F (email: James.Freericks@georgetown.edu)
\end{addendum}

\bibliography{Collection}

\end{document}